\documentclass{article} 
\usepackage[preprint]{colm2026_conference}

\usepackage{microtype}
\usepackage{hyperref}
\usepackage{url}
\usepackage{booktabs}
\usepackage{amsmath,amssymb}
\usepackage{multirow}
\usepackage{graphicx}
\usepackage{subcaption}
\usepackage{lineno}
\usepackage{newtxmath}

\definecolor{darkblue}{rgb}{0, 0, 0.5}
\hypersetup{colorlinks=true, citecolor=darkblue, linkcolor=darkblue, urlcolor=darkblue}

\title{Emergence of Reputation-Based Cooperation in LLM Agents}

 \author{Kazuya Horibe \\
   RIKEN \\
   \texttt{kazuya.horibe@riken.jp} \\
   \And
   Kenji Itao \\
   RIKEN  \\
   Tohoku University\\
   \And
   Wataru Toyokawa \\
   RIKEN}

\begin{document}

\ifcolmsubmission
\fi

\maketitle
\pagestyle{plain}

\begin{abstract}
Can cooperation among large language model (LLM) agents be evolutionarily stable against free-rider invasion?
We study an indirect reciprocity donation game where LLM agents observe behavioral traces and donate on a continuous scale.
Strategies, represented as natural language prompts, evolve through cultural transmission across generations.
Across four LLM backends, robustness to free-rider invasion varies by more than an order of magnitude.
The strongest predictor of this robustness is opponent endowment sensitivity, the degree to which agents discriminate between cooperative and uncooperative opponents, operationalizing the classical Image Scoring mechanism.
By contrast, adherence to the Leading-Eight L1 norm does not predict robustness.
Robustness depends on defector exclusion: while both cooperator reward and defector punishment vary across models, only the stringency of defector exclusion predicts resistance to free-rider invasion.
These findings reveal that LLM agents are confined to Image Scoring-like discrimination and fail to develop the more robust Leading-Eight norms, highlighting a fundamental vulnerability in culturally evolved LLM cooperation and motivating bottom-up approaches to norm construction.

\end{abstract}

\section{Introduction}

The rapid proliferation of large language models (LLMs) is reshaping how humans and artificial agents interact \citep{burton2024large, tsvetkova2024new}.
LLMs can simulate human-like responses across a wide range of social and economic scenarios \citep{akata2025playing, brookins2023playing, fan2024can, guo2024economics, horton2023large}.
As they increasingly participate in real-world contexts, ranging from customer service to collaborative work, their behaviors directly influence human behavior, creating feedback loops that shape future training data \citep{perez2024cultural}.
This ``co-evolution'' between human and AI behavior makes understanding LLM social dynamics not only an academic exercise but also a practical concern for designing robust human-AI hybrid societies \citep{dafoe2020open, conitzer2023foundations}.

A central question in this domain is whether LLM agents can sustain cooperation when self-interested exploitation is possible.
Cooperation among self-interested agents is a foundational problem across biology, social sciences, and artificial intelligence \citep{hamilton1964genetical, trivers1971evolution, axelrod1981evolution, nowak2006five}.
Five general mechanisms promote the evolution of cooperation \citep{nowak2006five}, and experimental work has confirmed that humans readily cooperate and punish free-riders even in anonymous one-shot settings \citep{fehr2002altruistic}.
Unlike direct reciprocity, which sustains cooperation in repeated dyadic interactions \citep{axelrod1980effective, trivers1971evolution}, indirect reciprocity operates when the beneficiary of cooperation is not the one who later reciprocates \citep{alexander1987biology, nowak2005evolution}. Cooperative acts build reputation, and cooperation is sustained when others preferentially help those with good reputations \citep{wedekind2000cooperation, milinski2002reputation, yoeli2013powering}.


Two theoretical frameworks address cooperation through indirect reciprocity, differing fundamentally in their information structure \citep{nowak2005evolution, okada2020review}.
Image Scoring \citep{nowak1998evolution} is a one-dimensional system where agents track observed behavior and cooperate with those whose image score exceeds a threshold.
This mechanism is intuitive and mirrors human gossip-based reputation systems \citep{wedekind2000cooperation}, but early models showed vulnerability to observation errors: occasional misperception of an action can cause cascading reputation loss, destabilizing cooperation \citep{leimar2001evolution, panchanathan2003tale}.
To address this limitation, \citet{ohtsuki2004should, ohtsuki2006leading} introduced the Leading Eight, eight evolutionarily stable norms that update reputations based on both the donor's action and the recipient's reputation, forming a two-dimensional system where agents track ``what did this person do, and to whom did they do it?''
This additional dimension enables more sophisticated discrimination: for example, the L1 norm \citep{fujimoto2023evolutionary} prescribes that refusing to help a person with a bad reputation should not damage one's own reputation, making the system more robust to errors.
Subsequent work has demonstrated that the complexity of social norms and the structure of past reputations critically affect the stability of cooperation \citep{santos2018social, hilbe2018indirect, schmid2021unified}.
However, the Leading Eight framework assumes that agents have access to explicit binary reputation labels (``good'' or ``bad''). In real-world settings, such labels are typically not directly observable. Instead, agents must infer reputations based on observed behavioral histories, which may be noisy or incomplete.

Despite these theoretical advances, most prior work has studied indirect reciprocity using agents with binary action spaces (cooperate or defect), publicly-shared reputation labels (``good'' or ``bad''), and the evolutionary assumption where behavioral strategies inherit directly through natural selection (the phenotypic gambit).
These simplifications facilitate analytical tractability but diverge from real-world settings in three critical ways.
First, human social interactions typically involve continuous decisions, not ``help or refuse'' but ``how much to help,'' ranging from token gestures to substantial sacrifices.
Continuous action spaces qualitatively change strategic dynamics: agents can implement graded responses, cooperating partially with uncertain partners and reserving full cooperation for those with strong reputations \citep{roberts1998development, wahl1999continuous, schmid2023quantitative}.

Second, humans and LLM agents do not always share reputation labels; instead, they must evaluate their opponents' goodness from behavioral observations, which are often incomplete and noisy \citep{hilbe2018indirect}.
This makes the system effectively Image Scoring-like (one-dimensional reputation inference) rather than Leading-Eight-like (two-dimensional reputation assignment).
Third, LLM populations evolve through cultural transmission of natural-language prompt texts \citep{perez2024cultural}.
Biases in generated text and the potentially nonlinear genotype--phenotype mapping (prompt text to behavioral strategy) may qualitatively alter evolutionary dynamics in ways the phenotypic gambit cannot predict.
Whether cultural transmission, which allows rapid behavioral adaptation \citep{boyd1985culture, mesoudi2011cultural, henrich2004cultural}, produces the same cooperation mechanisms as genetic evolution remains an open question.

LLM agents, equipped with natural language reasoning and trained on human social data, provide a new testbed for examining whether reputation-based cooperation emerges spontaneously in continuous action spaces, and how such strategies relate to theoretical frameworks like Image Scoring and the Leading Eight.



A growing literature examines LLM behavior in strategic interactions \citep{akata2025playing, brookins2023playing, fan2024can, guo2024economics, horton2023large}, generally focusing on direct reciprocity.
These studies reveal that LLMs exhibit cooperative tendencies that vary systematically across models, contexts, and game structures \citep{mei2024turing, lore2024strategic, fontana2025nicer}.
\citet{perez2024cultural} demonstrated that LLM populations exhibit cultural evolution dynamics analogous to human populations, with emergent attractors and cumulative changes across generations.
Most directly relevant, \citet{vallinder2024cultural} studied donation games among LLM agents and found that cooperative behavior evolves in Claude but not in Gemini or GPT-4o. However, two important questions remain unaddressed: the mechanisms driving this model-dependent variation are poorly understood, and whether evolved cooperation is robust against invasion by unconditional defectors (ALLD), that is, whether cooperative populations can resist exploitation by free-riders, was not tested.

This robustness question is not merely academic. As LLM agents are increasingly deployed in real-world collaborative settings alongside humans, naive cooperation that collapses under adversarial exploitation poses a practical risk. In human--AI hybrid societies, agents must not only cooperate with well-intentioned partners but also resist manipulation by malicious actors \citep{dafoe2020open}. Understanding whether culturally evolved cooperation in LLM populations is robust to free-rider invasion, and identifying the mechanisms that confer such robustness, is therefore essential for the safe deployment of autonomous agents.

In this paper, we study an indirect reciprocity donation game in which LLM agents observe behavioral traces of potential recipients (a chain of the last three interactions) and decide how much to donate on a continuous scale (0--100\% of their endowment).
Strategies are defined as generated texts, and evolve through cultural transmission: offspring inherit their parent's strategy and modify it based on the parent's performance rank, using the LLM itself as the mutation operator.
We tested four LLM backends across ten generations of twelve agents each, measuring both population-level resource accumulation and robustness to free-rider invasion.

\section{Methods}
Our experimental framework builds on \citet{vallinder2024cultural}, who studied cultural evolution in LLM donation games with top-$k$ truncation selection (the top 50\% of agents survive and each produces one offspring).
We modify two aspects: we replace truncation selection with fitness-proportional (roulette wheel) selection, where each agent's probability of being selected as a parent is proportional to its accumulated resources, and we redesign the prompts given to agents for both gameplay and strategy evolution (full prompt texts in Appendix~\ref{app:prompts}).

$N = 12$ LLM agents play an indirect reciprocity donation game across $G = 10$ generations.
Each agent decides what fraction (0--100\%) of their endowment to donate to recipients, with a 2:1 benefit-to-cost ratio ($b/c = 2$).
Agents observe three-round behavioral traces (e.g., ``recipient A donated $x$\% to B, B donated $y$\% to C'') rather than explicit reputation labels, making this an observation-based system.
Each agent's strategy is a natural language text (up to 350 words).
In generation 1, the LLM generates an initial strategy given the game rules (Appendix~\ref{app:prompts}, Prompt~1).
At each subsequent generation, roulette wheel selection chooses $N$ parents (with replacement), and each offspring inherits its parent's strategy and generates a modified version based on the parent's performance rank (Appendix~\ref{app:prompts}, Prompt~2).
The LLM thus serves as both the decision-maker during gameplay (Appendix~\ref{app:prompts}, Prompt~3) and the mutation operator during evolution.
We test four LLM backends: Claude 3.5 Sonnet, Gemini 1.5/2.0/2.5 Flash, with 10 independent runs per model.

To measure robustness, we conduct bot invasion experiments.
For each evolved strategy, we populate a 12-agent game with 11 copies of that strategy and 1 free-rider bot (always donates 0\%), then run a single-generation round-robin game without strategy evolution.
A strategy is classified as robust if the evolved agents' average score exceeds the bot's score.
We test approximately 100 strategies per model (10 randomly sampled per generation $\times$ 10 generations); robustness is the percentage classified as robust.

We quantify strategies via $11 \times 11$ cooperation matrices $M$, where $M_{ij}$ represents the agent's donation when $x_B = i \times 10\%$ (B's donation in round $t{-}2$) and $x_A = j \times 10\%$ (A's donation in round $t{-}1$).
Both axes represent observed opponent behavioral signals: $x_A$ is the recipient's own recent cooperation, while $x_B$ is the cooperation level of the recipient's prior interaction partner, consistent with Image Scoring assessment.
From these matrices, we extract features including: opponent endowment sensitivity = right\_edge (cooperation when $x_A = 60$--$100\%$) $-$ left\_edge (cooperation when $x_A = 0$--$40\%$); corner values $\alpha, \beta, \gamma, \delta$; L1 prevalence; monotonicity (Spearman $\rho$); and entropy.
Following \citet{ohtsuki2004should}, we compute four corner values from $5 \times 5$ sub-regions of $M$:
$\alpha$ = mean($M$[high $x_B$, low $x_A$]),
$\beta$ = mean($M$[high $x_B$, high $x_A$]),
$\gamma$ = mean($M$[low $x_B$, low $x_A$]),
$\delta$ = mean($M$[low $x_B$, high $x_A$]),
where ``high'' denotes indices 6--10 and ``low'' denotes indices 0--4.
From these we define:
\begin{equation}
\text{Opponent Endowment Sensitivity (OES)} = \underbrace{\frac{1}{55} \sum_{i=0}^{10} \sum_{j=6}^{10} M_{ij}}_{\text{right\_edge}} - \underbrace{\frac{1}{55} \sum_{i=0}^{10} \sum_{j=0}^{4} M_{ij}}_{\text{left\_edge}}
\end{equation}
Discrimination strength is $\beta - \alpha$.
An agent is classified as strongly monotonic if the mean Spearman $\rho$ between column index and cooperation value across all rows exceeds 0.5.
The L1 condition \citep{ohtsuki2004should} requires $\alpha < \gamma < \delta \leq \beta$.
Full definitions of entropy and additional details appear in Appendix~\ref{app:methods}.

\section{Results}

\subsection{Evolutionary dynamics and model-dependent robustness}

\begin{figure}[t]
\centering
\includegraphics[width=1.0\textwidth]{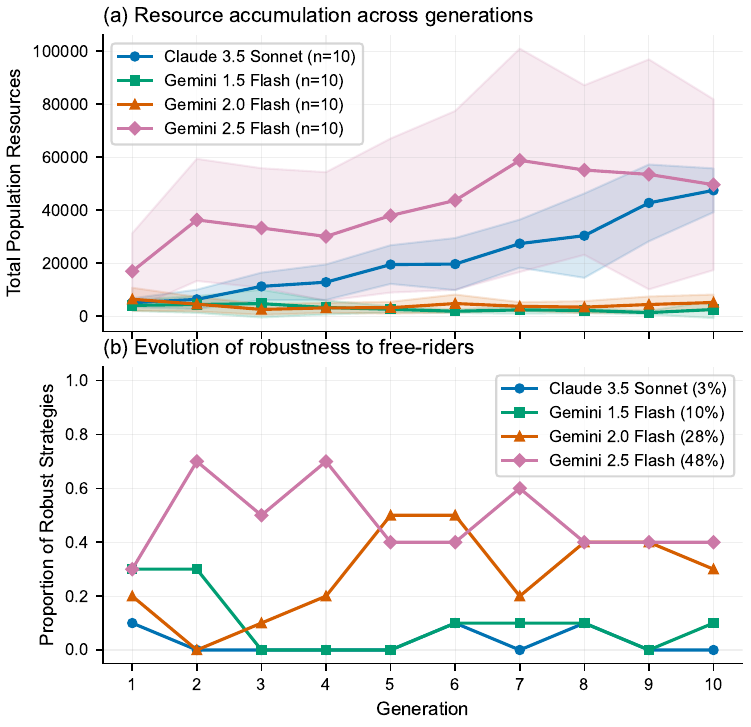}
\caption{Evolution of cooperation and robustness across generations (x-axis shared). \textbf{(a)} Mean total resources per generation for each LLM backend. Shaded regions indicate standard deviation across 10 independent runs. Claude 3.5 Sonnet and Gemini 2.5 Flash show substantial increases, while Gemini 1.5/2.0 Flash show modest or no growth. \textbf{(b)} Proportion of strategies from each generation that successfully expelled a free-rider bot (always donates 0\%). Final robustness: Gemini 2.5 Flash (48\%), Gemini 2.0 Flash (28\%), Gemini 1.5 Flash (10\%), Claude 3.5 Sonnet (3\%).}
\label{fig:evolution}
\end{figure}

We begin by examining how cooperation and robustness to free-rider invasion evolve across generations (Figure~\ref{fig:evolution}).

\paragraph{Resource accumulation varies widely across models.}
Claude~3.5 Sonnet and Gemini~2.5 Flash show substantial increases in mean total resources over generations (Figure~\ref{fig:evolution}a), indicating the evolution of cooperative strategies that benefit the population as a whole.
Consistent with prior findings \citep{vallinder2024cultural}, these models successfully develop and maintain strategies that enable mutual cooperation, resulting in resource growth that exceeds the initial endowment by factors of 2--3 by generation 10.
In contrast, Gemini~1.5 Flash and Gemini~2.0 Flash show modest or no increase in resources, suggesting that their evolved strategies fail to sustain high levels of cooperation across the population.
The variance across independent runs (shaded regions in Figure~\ref{fig:evolution}a) indicates that evolutionary outcomes are relatively stable within each model, though Gemini~2.5 Flash exhibits higher variability, possibly reflecting a richer set of viable cooperative strategies.

\paragraph{Robustness to free-rider invasion varies by an order of magnitude.}
Despite achieving similar resource growth, models differ substantially in robustness to free-rider invasion (Figure~\ref{fig:evolution}b).
Gemini~2.5 Flash achieves the highest robustness at 48\%, meaning that nearly half of its evolved strategies can successfully resist exploitation when a free-rider infiltrates the population.
Gemini~2.0 Flash follows with 28\% robustness, Gemini~1.5 Flash with 10\%, and Claude~3.5 Sonnet with only 3\%.
The most robust model is thus 16 times more robust than the least robust, a difference spanning 45 percentage points.

The temporal dynamics of robustness emergence (Figure~\ref{fig:evolution}b) reveal distinct evolutionary pathways.
Gemini~2.5 Flash develops robust strategies early, with robustness peaking at 70\% in generation 2 before stabilizing around 40--48\% in later generations.
This early emergence suggests that the model's initial strategy generation already biases toward defector exclusion mechanisms, which are then refined through cultural transmission.
Gemini~2.0 Flash shows a different pattern: robustness fluctuates considerably across generations, reaching 50\% in generations 5--6 before declining to 30\% by generation 10.
This volatility may reflect competing evolutionary pressures between cooperation and exploitation within the population.
Claude~3.5 Sonnet, by contrast, shows minimal robustness throughout evolution, never exceeding 10\% in any generation.
This persistent vulnerability indicates that the model's evolved strategies systematically fail to exclude free-riders, despite successfully cooperating among themselves.

\subsection{Mechanisms of robustness: continuous discrimination}
\label{sec:mechanisms}

All four models develop threshold-based discrimination without explicit instruction to do so.
Across 400 agents spanning four models and ten generations (Appendix~\ref{app:profiles}), 98.8\% exhibit strongly monotonic strategies (Spearman $\rho > 0.5$ between opponent cooperation level and agent cooperation level), with model-specific rates ranging from 97.0\% to 100.0\% and mean correlations from 0.931 to 0.970.
This near-universal emergence of monotonic strategies confirms that Image Scoring-like discrimination \citep{nowak1998evolution} is a robust emergent property of LLM agents.
Yet if all models discriminate, why does robustness to free-riders vary 16-fold?

\paragraph{Discrimination strength, not its presence, predicts robustness.}
The answer lies not in whether agents discriminate, but in how strongly they do so.
We regressed model-level robustness against model-level mean strategy features (Table~\ref{tab:regression}).

\begin{table}[h]
\centering
\caption{Single-feature regression predicting robustness (\%) from generation 10 model-level mean strategy features ($n = 4$ models). $^{*}$: $p < 0.05$; $^{\dagger}$: $p < 0.10$.}
\label{tab:regression}
\begin{tabular}{lcccc}
\toprule
\textbf{Feature} & $\boldsymbol{R^2}$ & \textbf{$p$-value} & \textbf{Slope} & \textbf{Sig.} \\
\midrule
Opponent endow.\ sensitivity & 0.934 & 0.033 & +1.62 & $^{*}$ \\
$\beta - \alpha$           & 0.898 & 0.053 & +1.73 & $^{\dagger}$ \\
Matrix entropy             & 0.788 & 0.112 & $-219.3$ & n.s. \\
Left edge                  & 0.786 & 0.113 & $-2.30$ & n.s. \\
L1 prevalence              & 0.671 & 0.181 & $-116.7$ & n.s. \\
Mean cooperation           & 0.046 & 0.785 & $-0.98$ & n.s. \\
\bottomrule
\end{tabular}
\end{table}

Opponent endowment sensitivity (Equation~1) explains 93.4\% of the variance in robustness ($R^2 = 0.934$, $p = 0.033$; Table~\ref{tab:regression}), making it the only significant predictor at $\alpha = 0.05$ with $n = 4$ models.
The closely related metric $\beta - \alpha$ ($R^2 = 0.898$, $p = 0.053$; $r > 0.99$ with opponent endowment sensitivity) corroborates this finding.
By contrast, mean cooperation level has virtually no predictive power ($R^2 = 0.046$), and L1 prevalence, the proportion of strategies satisfying the Leading-Eight L1 condition \citep{ohtsuki2004should}, does not significantly predict robustness ($R^2 = 0.671$, $p = 0.181$).
The L1 failure is expected: the L1 norm assumes access to explicit binary reputation labels, whereas our agents infer cooperativeness from behavioral traces, making this effectively an Image Scoring system where one-dimensional discrimination is the relevant mechanism.

\paragraph{Defector exclusion predicts robustness.}
To understand \emph{why} discrimination strength predicts robustness, we decompose opponent endowment sensitivity into its two components (Table~\ref{tab:sensitivity}).

\begin{table}[h]
\centering
\caption{Decomposition of opponent endowment sensitivity (generation 10 means, $n = 10$ agents per model). Left edge = cooperation when $x_A = 0$--$40\%$ (A's donation in round $t{-}1$); right edge = cooperation when $x_A = 60$--$100\%$.}
\label{tab:sensitivity}
\begin{tabular}{lcccc}
\toprule
\textbf{Model} & \textbf{Left Edge} & \textbf{Right Edge} & \textbf{Opp.\ Endow.\ Sens.} & \textbf{Robustness (\%)} \\
\midrule
Gemini 2.5 Flash    & 22.4 & 83.5 & 61.1 & 48.0 \\
Gemini 2.0 Flash    & 22.6 & 73.6 & 51.0 & 28.0 \\
Claude 3.5 Sonnet   & 38.7 & 76.9 & 38.2 & 3.0 \\
Gemini 1.5 Flash    & 30.8 & 65.9 & 35.1 & 10.0 \\
\bottomrule
\end{tabular}
\end{table}

The right edge (cooperation with high-cooperation opponents) varies across models (65.9--83.5\%, a span of 17.6 percentage points), though all models reward cooperators at moderately high levels.
The left edge (cooperation with low-cooperation opponents) spans a similar range (22.4--38.7\%, a span of 16.3 percentage points).
Crucially, the absolute level of cooperation with low-cooperation opponents determines robustness: models with left edges below 25\% (Gemini 2.0 and 2.5 Flash) achieve substantially higher robustness than those above 30\% (Claude 3.5 Sonnet and Gemini 1.5 Flash).
The mechanism is straightforward: a free-rider that always donates 0\% will be identified as a low-cooperation opponent.
Claude~3.5 Sonnet, with a left edge of 38.7\%, donates nearly two-fifths of its resources even to such opponents, allowing free-riders to accumulate 7.74 units per round ($0.387 \times 10 \times 2$, accounting for the 2:1 benefit-to-cost ratio), which can exceed the net gains of cooperators.
Gemini~2.0 and 2.5 Flash maintain left edges near 22\%, limiting free-rider accumulation to approximately 4.5 units per round, insufficient to outcompete cooperators whose mutual cooperation yields higher returns.

\paragraph{Strategy texts confirm the quantitative pattern.}
Table~\ref{tab:strategies} and Figure~\ref{fig:heatmaps} (Appendix~\ref{app:heatmaps}) show what discrimination strength looks like in practice.
Robust strategies (Gemini 2.5 Flash) include explicit punishment thresholds: ``if the recipient (A) donated less than 20\%, donate 0\%.''
This threshold produces the steep left-to-right gradient visible in the heatmaps: near-zero cooperation with low-cooperation opponents, high cooperation with high-cooperation opponents.
Non-robust strategies (Claude 3.5 Sonnet) instead maintain cooperation floors: ``but never less than 35\%,'' producing the uniformly elevated heatmaps that allow free-riders to persist.
A multiple regression incorporating $\beta - \alpha$ and left edge achieves $R^2 = 0.961$ (MAE = 3.23 pp; note that with $n = 4$ and two predictors, residual degrees of freedom are minimal), suggesting that discrimination strength and defector exclusion jointly account for robustness.

\begin{table}[h]
\centering
\caption{Example evolved strategies from generation 10. \textbf{Robust} strategies explicitly exclude low-cooperation opponents; \textbf{Non-robust} strategies maintain positive cooperation even toward defectors.}
\label{tab:strategies}
\small
\begin{tabular}{p{2cm}p{10cm}}
\toprule
\textbf{Type} & \textbf{Strategy Text (abbreviated)} \\
\midrule
\textbf{Robust} (Gemini 2.5) & ``Donate 100\% in the first round; thereafter, if the recipient (A) donated 80\% or more, donate 100\%; \textbf{if A donated less than 20\%, donate 0\%}; if A donated between 20\% and 79\%, determine a base increase based on B's donation...'' \\
\midrule
\textbf{Non-robust} (Claude 3.5) & ``Donate 50\% in the first round; in subsequent rounds, donate an amount equal to the average of A's last donation and the average of B and C's donations, plus a base 25\% that increases by 1\% each round up to 40\%, \textbf{but never less than 35\%}...'' \\
\bottomrule
\end{tabular}
\end{table}

\section{Discussion}

\subsection{From classical Image Scoring to continuous discrimination}

Our results are consistent with the Image Scoring mechanism \citep{nowak1998evolution} in a cognitively rich setting with continuous actions.
Four predictions from Image Scoring theory are supported:
(1) threshold-based strategies evolve (98.8\% of agents show monotonic strategies);
(2) agents discriminate based on observed cooperation (opponent endowment sensitivity predicts robustness with $R^2 = 0.934$);
(3) discriminators exclude defectors (high sensitivity correlates with high robustness, $r = 0.967$); and
(4) no explicit reputation system is needed (observation-based inference suffices).
These findings extend Image Scoring from the domain of simple, genetically evolved agents to LLM agents with natural language reasoning and cultural transmission.
The fact that LLMs develop these strategies without explicit instruction to discriminate suggests that cooperative discrimination is a robust emergent property of agents trained on human social data.

\paragraph{LLM agents are confined to Image Scoring.}
L1 prevalence does not predict robustness ($R^2 = 0.671$, $p = 0.181$), and only 9--15\% of strategies across all models satisfy the L1 condition.
This means that LLM agents, despite their sophisticated language capabilities, fail to construct the more robust Leading-Eight norms that theory predicts are necessary for cooperation to withstand observation errors \citep{ohtsuki2006leading}.
Instead, they are confined to Image Scoring, a mechanism known to be vulnerable to noise and misperception \citep{leimar2001evolution, panchanathan2003tale}.
Even the most robust model (Gemini~2.5 Flash) achieves only 48\% robustness against a simple unconditional defector, suggesting that Image Scoring alone is insufficient to guarantee evolutionary stability in realistic settings where observation errors are inevitable.

\paragraph{Generalizing Image Scoring to continuous action spaces.}
These findings suggest that Image Scoring naturally extends to continuous action spaces.
In the classical binary setting, Image Scoring prescribes a threshold rule: cooperate if the opponent's reputation exceeds a cutoff, defect otherwise \citep{nowak1998evolution}.
Our agents implement a continuous analogue: a graded cooperation function where donation increases with observed opponent cooperation.
The critical insight is that monotonicity alone is insufficient: nearly all agents (98.8\%) develop monotonic strategies, yet robustness varies 16-fold.
What distinguishes robust from non-robust strategies is the \emph{steepness} of the discrimination gradient, captured by opponent endowment sensitivity.
Claude~3.5 Sonnet illustrates this: despite near-universal monotonicity (99\% of strategies), its shallow gradient (OES = 38.2) fails to exclude defectors, yielding only 3\% robustness.
This suggests that the relevant dimension for cooperation stability in continuous spaces is not whether agents discriminate, but how sharply they do so, representing a quantitative rather than qualitative criterion.

\subsection{Mechanisms of robustness and cultural evolution}

\paragraph{Defector exclusion as the critical axis of variation.}
As shown in Section~\ref{sec:mechanisms}, the critical differentiator of robustness is not how much agents reward cooperators, but how stringently they exclude defectors.
This suggests that the evolutionary challenge for cooperative LLM populations is calibrating the threshold at which cooperation is withheld from suspected defectors.

\paragraph{Cultural evolution refines rather than discovers discrimination.}
Unlike genetic evolution, where discrimination mechanisms must evolve de novo from random strategies, LLMs begin with strong priors for cooperative discrimination, likely inherited from pre-training on human social interaction data.
Discrimination strength ($\beta - \alpha$) is already high in generation~1 (ranging from 39 to 59 across models; Table~\ref{tab:profiles}) and remains largely stable through generation~10, with three of four models changing by less than 6\% (Gemini~1.5 Flash decreases by 19\%, the sole exception).
This stability suggests that cultural evolution in LLM populations operates primarily through refinement and recombination of existing strategies rather than discovery of qualitatively new behavioral patterns.
The initial diversity in discrimination strength persists throughout evolution (Table~\ref{tab:profiles}, gen.\ 1 vs.\ gen.\ 10 rows), indicating that model-specific priors, established during pre-training, constrain the space of strategies accessible through cultural transmission.
This contrasts sharply with genetic evolution studies, where cooperation must bootstrap from initially uncooperative populations \citep{ohtsuki2006leading}.
This suggests that multi-agent AI systems may not need to evolve cooperation from scratch; rather, the challenge lies in steering inherited cooperation tendencies toward robustness by strengthening defector exclusion mechanisms.

\subsection{Vulnerability of LLM cooperation and implications for norm construction}

Our findings reveal a fundamental tension in LLM cooperation.
On one hand, some evolved strategies successfully resist invasion by unconditional defectors (ALLD): up to 48\% of Gemini~2.5 Flash strategies outcompete a free-rider bot.
On the other hand, LLM agents universally fail to develop Leading-Eight norms, which are theoretically necessary for cooperation to remain stable under observation errors \citep{ohtsuki2006leading, santos2018social}.
This creates a fragile form of cooperation: while agents can exclude obvious defectors through Image Scoring, they lack the sophisticated norm structure needed to handle the ambiguity that arises when cooperative acts are occasionally misperceived.

This vulnerability has practical implications for the deployment of LLM agents in human--AI hybrid societies.
Current approaches to aligning LLM behavior rely primarily on top-down, human-in-the-loop methods, such as reinforcement learning from human feedback (RLHF) and constitutional AI, that instill cooperative norms exogenously \citep{ouyang2022training, bai2022constitutional}.
Our results suggest that such top-down norm instillation may be insufficient: even when agents are trained on vast corpora of cooperative human behavior, they fail to develop the second-order discrimination (conditioning on the recipient's reputation, not just their behavior) that characterizes robust norms.
An alternative approach would be to allow agents to construct norms \emph{bottom-up}, through direct experience with exploitation and norm violation.
If agents could learn from repeated encounters with free-riders, rather than merely inheriting cooperative priors from pre-training data, they might develop the context-sensitive discrimination that the Leading Eight prescribe, building norms that are robust to both unconditional defection and observation errors.
Whether such bottom-up norm construction is achievable in LLM populations remains an open question for future work.

\subsection{Limitations}

Several limitations should be noted.
First, statistical analysis is at the model level ($n = 4$), limiting power despite large effect sizes ($R^2 = 0.93$).
Second, we test one population size ($N = 12$); the relationship between size and robustness remains unknown.
Third, bot invasion uses a single, non-adaptive free-rider; real scenarios may involve multiple or adaptive exploiters.
Fourth, we do not analyze natural language strategy content.
Fifth, strategy matrices are computed from synthetic scenarios, which may not perfectly reflect dynamic game behavior.

Despite these limitations, our findings provide initial theoretical and empirical evidence that Image Scoring is the relevant mechanism in this setting, and that opponent endowment sensitivity serves as a tractable predictor of robustness.

\section{Conclusion}
Our central finding is that LLM agents, despite sophisticated language capabilities, are confined to Image Scoring-like discrimination and fail to develop the Leading-Eight norms that theory identifies as necessary for robust cooperation under observation errors.
Opponent endowment sensitivity provides a tractable diagnostic for predicting free-rider resistance, but even the strongest discriminators achieve only 48\% robustness, revealing a fundamental vulnerability in culturally evolved LLM cooperation.
The practical challenge for multi-agent AI systems is therefore twofold: LLMs inherit cooperative priors from pre-training, but these priors produce only first-order discrimination (Image Scoring), leaving populations vulnerable to the cascading reputation loss that observation errors can trigger.

Future work should explore whether bottom-up norm construction, allowing agents to learn discrimination rules from direct experience with exploitation rather than inheriting them from training data, can produce the second-order norms (Leading Eight) that confer robustness to both defection and noise.
Additionally, testing stability in hybrid human--LLM populations and investigating how training procedures shape the space of accessible norms remain important open questions.

\bibliography{colm2026_conference}
\bibliographystyle{colm2026_conference}

\appendix

\section{LLM Usage Statement}
In the process of developing this work, we utilized LLMs as a reviewer tool to improve the quality of the manuscript, including grammar correction and rewriting. In addition, we employed LLMs for data visualization.

\section{Detailed methods}
\label{app:methods}

\subsection{Game structure}

Our framework builds on \citet{vallinder2024cultural} with the following modifications:

\begin{itemize}
\item Selection mechanism: \citet{vallinder2024cultural} use top-$k$ truncation selection (top 50\% survive, each produces one offspring). We use fitness-proportional roulette wheel selection, where each agent's probability of being selected as a parent is $p_i = f_i / \sum_j f_j$, with $f_i$ being agent $i$'s total accumulated resources. Parents are sampled with replacement to produce $N$ offspring.
\item Prompts: We redesign the system prompt, strategy generation prompt, and donation decision prompt (see Appendix~\ref{app:prompts}).
\item Cultural transmission mode: At each generation boundary, each offspring receives its parent's strategy text and performance rank, then generates a modified strategy. The LLM serves as both the inheritor and the mutation operator, creating variation while preserving successful elements.
\end{itemize}

The game structure is otherwise identical. A population of $N = 12$ agents plays across $G = 10$ generations.
Within each generation, agents are paired using a bipartite round-robin scheme played twice with reassigned roles, yielding $2 \times 2 \times \frac{N}{2} \times \frac{N}{2} = 144$ donation events per generation for $N = 12$.
Each agent begins with 10 units of resources. If a donor gives $x$ units, the donor loses $x$ and the recipient gains $2x$ ($b/c = 2$).
Strategy generation uses temperature 0.8; gameplay decisions use temperature 0.3.

\subsection{Information structure}

Agents do not observe explicit reputation labels. When paired with recipient $A$ in round $t$, the donor observes a three-round behavioral chain: (1) in round $t{-}1$, $A$ donated $x_A$\% to $B$; (2) in round $t{-}2$, $B$ donated $x_B$\% to $C$; (3) in round $t{-}3$, $C$ donated $x_C$\% to $D$.

\subsection{Bot invasion experiments}

We test the robustness of individual evolved strategies against a free-rider bot.
For each model, we randomly sample up to 10 strategies per generation across 10 generations, yielding approximately 100 strategies per model.
The sampled strategy identities are recorded via cryptographic hashes, ensuring that the same set is used consistently across all downstream analyses (matrix extraction, feature computation, and bot invasion tests).
For each strategy to be tested:
\begin{enumerate}
\item Create a population of 12 agents: 11 agents all using the same evolved strategy, plus 1 free-rider bot whose strategy is ``Always donate 0 units regardless of the situation.''
\item Run a single-generation round-robin game (no strategy evolution; \texttt{enable\_mutation=False}).
\item Compute the average final score of the 11 evolved agents and the final score of the bot.
\item Classify the strategy as robust if the evolved agents' average score exceeds the bot's score.
\end{enumerate}
Robustness is the percentage of tested strategies classified as robust.
By using 11 copies of the same strategy rather than a mixture of different strategies, this test isolates whether a given strategy can resist exploitation independently of population composition.
Bot invasion experiments use the Gemini API's default temperature for all gameplay LLM calls, rather than the explicit temperature 0.3 used during evolution.
Because the single-generation test involves no strategy generation phase, only gameplay decisions are affected; the higher default temperature introduces additional stochasticity in donation decisions, providing a conservative robustness test.

\subsection{Strategy matrix extraction}

We construct an $11 \times 11$ cooperation matrix $M$ for each agent by running 121 synthetic scenarios varying $x_B$ (B's donation in round $t{-}2$; row $i$, where $x_B = i \times 10\%$) and $x_A$ (A's donation in round $t{-}1$; column $j$, where $x_A = j \times 10\%$).

Following \citet{ohtsuki2004should}, we compute four corner values from $5 \times 5$ sub-regions:
$\alpha$ = mean($M$[high $x_B$, low $x_A$]),
$\beta$ = mean($M$[high $x_B$, high $x_A$]),
$\gamma$ = mean($M$[low $x_B$, low $x_A$]),
$\delta$ = mean($M$[low $x_B$, high $x_A$]).
L1 condition: $\alpha < \gamma < \delta \leq \beta$.

We define the following features from the cooperation matrix $M$:

\paragraph{Opponent endowment sensitivity.}
As defined in Section~2 (Equation~1), OES = right\_edge $-$ left\_edge, where left\_edge averages over columns $x_A = 0$--$40\%$ and right\_edge over $x_A = 60$--$100\%$.

\paragraph{Discrimination strength.}
As defined in Section~2, $\beta - \alpha$ measures the difference in cooperation between high-cooperation and low-cooperation opponents when the opponent's prior partner (B) showed high cooperation.

\paragraph{Matrix entropy.}
\begin{equation}
H(M) = -\sum_{i=0}^{10}\sum_{j=0}^{10} \tilde{M}_{ij} \ln \tilde{M}_{ij}, \quad \tilde{M}_{ij} = \frac{M_{ij}}{\sum_{k,l} M_{kl}}
\end{equation}

\paragraph{Monotonicity.}
We compute the Spearman rank correlation $\rho_S$ between column index ($x_A$ level) and cooperation value for each row $i$ of $M$, then average over all rows:
\begin{equation}
\rho_{\text{mono}} = \frac{1}{11} \sum_{i=0}^{10} \rho_S\bigl((0,\ldots,10),\; (M_{i,0},\ldots,M_{i,10})\bigr)
\end{equation}
An agent is classified as strongly monotonic if $\rho_{\text{mono}} > 0.5$.

\paragraph{L1 prevalence.} The proportion of agents satisfying $\alpha < \gamma < \delta \leq \beta$.

\section{Detailed model profiles}
\label{app:profiles}

\begin{table}[h]
\centering
\caption{Strategy profile summary for each LLM backend. Values represent model-level means across 100 agents.}
\label{tab:profiles}
\begin{tabular}{lcccc}
\toprule
\textbf{Feature} & \textbf{Gemini 2.5} & \textbf{Gemini 2.0} & \textbf{Claude 3.5} & \textbf{Gemini 1.5} \\
\midrule
Robustness (\%)         & 48.0 & 28.0 & 3.0  & 10.0 \\
Opp.\ endow.\ sensitivity & 61.2 & 51.6 & 40.8 & 40.6 \\
$\beta - \alpha$ (all gen.)  & 61.3 & 51.6 & 42.2 & 41.3 \\
$\beta - \alpha$ (gen.\ 1)  & 59.2 & 49.4 & 38.7 & 44.4 \\
$\beta - \alpha$ (gen.\ 10) & 60.7 & 50.8 & 40.7 & 35.9 \\
Left edge               & 21.7 & 24.7 & 35.2 & 30.6 \\
Right edge              & 82.9 & 76.3 & 75.9 & 71.2 \\
$\alpha$                & 23.5 & 26.6 & 37.7 & 32.7 \\
$\beta$                 & 84.8 & 78.2 & 79.9 & 74.0 \\
Monotonicity ($\rho$)   & 0.970 & 0.938 & 0.941 & 0.931 \\
Strongly monotonic (\%) & 100.0 & 97.0 & 99.0 & 99.0 \\
Entropy (normalized)    & 0.851 & 0.770 & 0.774 & 0.742 \\
Unique levels           & 19.5 & 14.5 & 18.4 & 18.2 \\
L1 prevalence (\%)      & 15.0 & 9.0  & 11.0 & 10.0 \\
\bottomrule
\end{tabular}
\end{table}

\section{Representative strategy heatmaps}
\label{app:heatmaps}

\begin{figure}[h]
\centering
\includegraphics[width=0.95\textwidth]{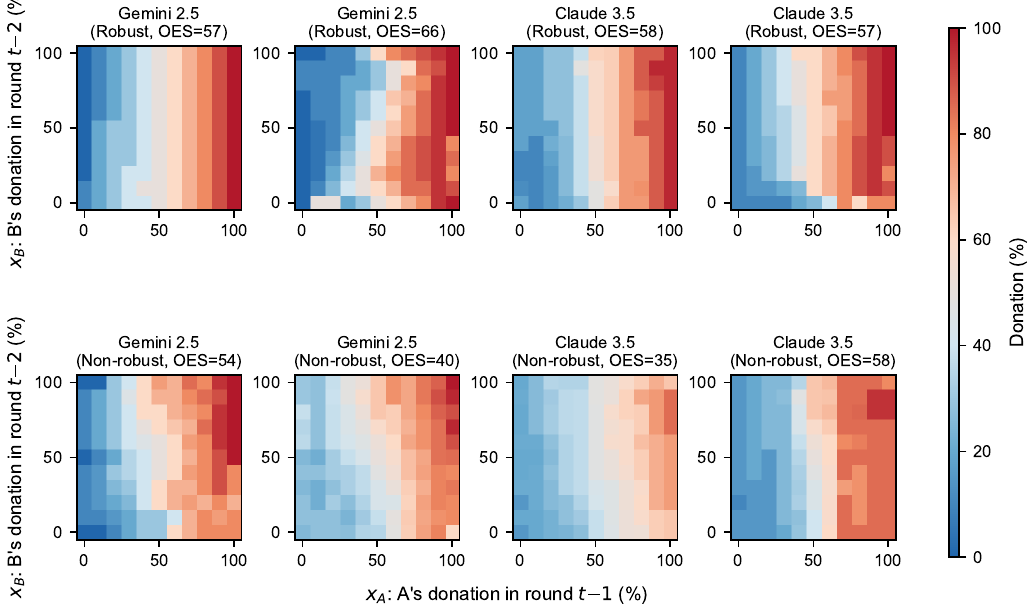}
\caption{Representative individual cooperation matrices (see Table~\ref{tab:strategies} for corresponding strategy texts). Columns = $x_A$ (A's donation in round $t{-}1$); rows = $x_B$ (B's donation in round $t{-}2$); cell color = agent's donation percentage. Top row: robust strategies (Gemini 2.5 Flash) that successfully expelled free-riders. Bottom row: non-robust strategies (Claude 3.5 Sonnet) that failed to exclude free-riders. OES = opponent endowment sensitivity. The steep gradient in robust strategies reflects explicit punishment thresholds; the uniform coloring in non-robust strategies reflects cooperation floors.}
\label{fig:heatmaps}
\end{figure}

Figure~\ref{fig:heatmaps} shows cooperation matrices from individual agents of Gemini 2.5 Flash and Claude 3.5 Sonnet, the most and least robust models respectively.
These heatmaps correspond to the types of natural language strategies shown in Table~\ref{tab:strategies}.
Robust strategies (top row, Gemini 2.5 Flash) exhibit a steep left-to-right gradient: cooperation is near zero when $x_A$ was low (left columns) and high when $x_A$ was high (right columns).
This pattern directly reflects the punishment threshold (``if A donated less than 20\%, donate 0\%'') in the strategy text.
Non-robust strategies (bottom row, Claude 3.5 Sonnet) show more uniform cooperation across the $x_A$ axis, with elevated values even in the left columns.
This reflects the cooperation floor (``but never less than 35\%'') that allows free-riders to accumulate resources passively.

\section{Prompts}
\label{app:prompts}

We provide the three core prompts used in our experiments. Variables in curly braces (\texttt{\{...\}}) are filled at runtime.

\paragraph{System prompt.}
The following system prompt is provided to all agents at the beginning of each game session:

\begin{quote}
\small
\label{prompt:system}
You are a game-playing agent. Your assistance will help understand multi-agent interactions.
You are playing the following game. Each player is given an initial endowment of \{initial\_endowment\} units of a resource.
In each round, you are randomly paired with another individual. One of you is a donor, and the other is a recipient.
The donor decides to give up some amount of the resource (between 0 and the total number of units they possess). The recipient receives \{cooperation\_gain\}$\times$ the number of units that the donor gave up.
If you were a donor in one round, you will be a recipient in the next round, and vice versa.

As a donor, you will receive limited information about the recipient and their recent interactions:
You will see what the recipient (let's call them A) did in the immediately preceding round only.
You will also see what A's partner from that round (let's call them B) did in their preceding round.
This chain continues for at most three rounds back.

Your goal is to maximize the number of units you have after the final round.
After the game has finished, the probability of producing offspring for the next generation is determined in proportion to the units each agent has after the final round (roulette selection method).
\end{quote}

\paragraph{Prompt 1: Initial strategy generation (generation 1).}
\label{prompt:init}

\begin{quote}
\small
Your name is \{agent\_name\}.
Based on the description of the game, create a strategy that you will follow in the game.

As a donor, you will receive limited information about the recipient and their recent interactions:
You will see what the recipient (let's call them A) did in the immediately preceding round only.
You will also see what A's partner from that round (let's call them B) did in their preceding round.
This chain continues for at most three rounds back.

In the first round, there will be no information about the recipient's previous behavior. Think carefully about how much to donate in that case.

Reflect briefly (do not reveal your internal reasoning). Output exactly two parts: first a brief 1--3 sentence reflection, then a single sentence starting with ``My strategy will be''.
\end{quote}

\paragraph{Prompt 2: Strategy evolution (generation $> 1$).}
\label{prompt:evolve}

\begin{quote}
\small
Your name is \{agent\_name\}.
How would you approach the game?
Modify this advice to create your own strategy. Here is information about your parent from the previous generation.

Your parent's strategy was: \{parent\_strategy\}

Your parent's rank in the previous generation was: \{parent\_rank\}/\{total\_agents\}

[Game observation rules repeated from system prompt.]

Before formulating your strategy, briefly think step by step about what would be a successful strategy in this game.
In particular, think about how you can improve on the surviving agents' strategies.
Then describe your strategy briefly without explanation in one sentence (up to 350 words) that starts: ``My strategy will be''.
\end{quote}

\paragraph{Prompt 3: Donation decision (gameplay).}
\label{prompt:donate}

\begin{quote}
\small
Your name is \{donor\_name\}.
As you will recall, here is the strategy you decided to follow: \{donor\_strategy\}

This is generation \{generation\}. It is now round \{round\}.
In this round, you have been paired with \{recipient\_name\}. They currently have \{recipient\_resources\} units of the valuable resource.

\{behavioral\_trace\}

You will take the role of donor. You currently have \{donor\_resources\} units of the valuable resource.
How many units do you give up?

Very briefly think step by step about how you apply your strategy in this situation and then provide your answer.

State your full response in the following form:\\
Justification (up to 100 words): [justification]\\
Answer: [number]
\end{quote}

\end{document}